\documentclass[conference]{IEEEtran}
\IEEEoverridecommandlockouts
\usepackage{graphicx}
\usepackage{tabularx}
\usepackage[justification=centering]{caption}
\usepackage{cite}
\usepackage{booktabs}
\usepackage{pdfpages}
\usepackage{subfigure}
\usepackage{atbegshi}% http://ctan.org/pkg/atbegshi
\AtBeginDocument{\AtBeginShipoutNext{\AtBeginShipoutDiscard}}

\ifCLASSINFOpdf

\else
\fi

\begin{document}
%
% paper title
% can use linebreaks \\ within to get better formatting as desired
\title{LAA-LTE and WiFi based Smart Grid Metering Infrastructure in 3.5 GHz Band}

% author names and affiliations
% use a multiple column layout for up to three different
% affiliations
%\author{\IEEEauthorblockN{Imtiaz Parvez, Nasidul Islam, Nadisanka Rupasinghe, Arif I. Sarwat, Ismail Guvenc }
%\IEEEauthorblockA{Department of Electrical and Computer Engineering\\
%Florida International University, Miami, FL, 33174\\
%Email: iparv001@fiu.edu, nisla004@fiu.edu, nadisanka@gmail.com, asarwat@fiu.edu, iguvenc@fiu.edu }

\author{\IEEEauthorblockN{Imtiaz Parvez${^{*}}$, Tanwir Khan${^{*}}$, Arif I. Sarwat${^{*1}}$ \thanks{${^{*1}}$Corresponding author}and Zakaria Parvez ${^{\dag}}$}
\IEEEauthorblockA{${^{*}}$Department of Electrical and Computer Engineering, Florida International University, Miami, FL 33174\\{${^{\dag}}$Department of Operation, Electricity Generation Company of Bangladesh.}\\ 
Email: {\tt \{iparv001,tkhan016,asarwat\}}@fiu.edu, zakaria.parvez@egcb.com.bd}}
\thanks{This research was supported in part by the U.S. National Science Foundation under the grant RIPS-1441223 and CAREER-0952977.}

%\author{\IEEEauthorblockN{Imtiaz Parvez${^{*}}$,  Tanwir Khan${^{*}}$, Arif I. Sarwat${^{*}}$, and Zakaria Parvez ${^{\dag}}$}
%	\IEEEauthorblockA{{${^{*}}$Department of Electrical and Computer Engineering, Florida International University, Miami, FL 33174.}\\ {${^{\dag}}$Electricity Generation Company of Bangladesh.}\\ {${^{\ddag}}$Department of Electrical and Electronic Engineering, Independent University, Dhaka, Bangladesh.}\\
%		Email: {\tt \{iparv001,tkhan016,asarwat\}@fiu.edu,zakaria.parvez@egcb.com.bd}}%
%	\thanks{This research was supported in part by the U.S. National Science Foundation under the grant RIPS-1441223 and CAREER-0952977.}
%}

% use for special paper notices
%\IEEEspecialpapernotice{(Invited Paper)}

%\IEEEoverridecommandlockouts
%\IEEEpubid{\makebox[\columnwidth]{978-1-5090-2246-5/16/\$31.00~
%\copyright2016
%IEEE \hfill} \hspace{\columnsep}\makebox[\columnwidth]{ }}

% make the title area
\maketitle

\begin{abstract}
%\boldmath

Advanced metering infrastructure (AMI) of smart grid requires bidirectional communication for transferring data to billing center, for which WiFi is an attractive choice. However, WiFi operates in the unlicensed bands and LTE needs to offload data in the same unlicensed band. Recent release of 3.5 GHz (also termed as citizen broadband radio service (CBRS))  can be an attractive shared band where LTE and WiFi can coexist. In our study, we propose a fixed duty cycled LTE-U and WiFi based smart grid metering infrastructure where smart meter uses WiFi and data collector (termed as Access Point (AP)) of smart meters  uses LTE for transferring data. We investigate the coexistence performance of LTE-WiFi in the 3.5~GHz band using a time division duplexing (TDD)-LTE confederated by WiFi along with FTP traffic model for system level simulation. The simulation results demonstrate a good neighboring coexistence between LTE and WiFi resulting a candidate AMI architecture for smart grid in the 3.5~GHz band.

\end{abstract}
% IEEEtran.cls defaults to using nonbold math in the Abstract.
% This preserves the distinction between vectors and scalars. However,
% if the conference you are submitting to favors bold math in the abstract,
% then you can use LaTeX's standard command \boldmath at the very start
% of the abstract to achieve this. Many IEEE journals/conferences frown on
% math in the abstract anyway.

% no keywords

% For peer review papers, you can put extra information on the coverd 
% page as needed:
% \ifCLASSOPTIONpeerreview
% \begin{center} \bfseries EDICS Category: 3-BBND \end{center}
% \fi
%
% For peerreview papers, this IEEEtran command inserts a page break and
% creates the second title. It will be ignored for other modes.
\IEEEpeerreviewmaketitle

\section{Introduction}
% no \IEEEPARstart

In energy and power sector, smart grid is the evolution of power system from one-way to two-way power system employing state-of-art approaches for intercommunication among its entities, and cutting edge techniques for delivering of electricity to the consumers with intensified efficiency and control mechanism~\cite{5635449, 7011570, DBLP:journals/corr/ParvezSPPK17}. The role of advanced metering infrastructure (AMI)  is very crucial  in the smart grid system as it forms the communication bridge between the consumer smart meters and metering data management service (MDMS) of smart grid for transferring consumption data using wireless communication~\cite{6939551, en9090691, 7462517}. The popular communication standard for AMI is WiFi, and Zigbee which utilizes unlicensed bands such as 900~MHz, 2.4~GHz, and 5~GHz~~\cite{v188}. In the unlicensed band, smart meters might need to share channel with other technologies such as LTE-unlicensed (LTE-U) and ZigBee.

Long-Term Evolution (LTE) is the standard for the nationwide broadband communications for the past several years \cite{5635449,DBLP:journals}. With the development in technology, LTE needs to support machine-to-machine (M2M) communication along with increased personal mobile communication. Furthermore, as the requirement of LTE data rate increases exponentially, scarcity of  spectrum becomes a critical issue. One solution that could be promising that  sharing of the spectrum among different wireless technologies, but this kind of coexistence mechanisms has its own implementation barriers. In particular, unlicensed spectrum can be utilized along with licensed band. To support this approach, there has also been research works by 3GPP standardization group on license assisted access (LAA) of LTE/ LTE-U in the unlicensed spectrum \cite{3GPP}.

\begin{figure*}[!tp]
    \centering
    \includegraphics[width=1\textwidth,scale=0.5]{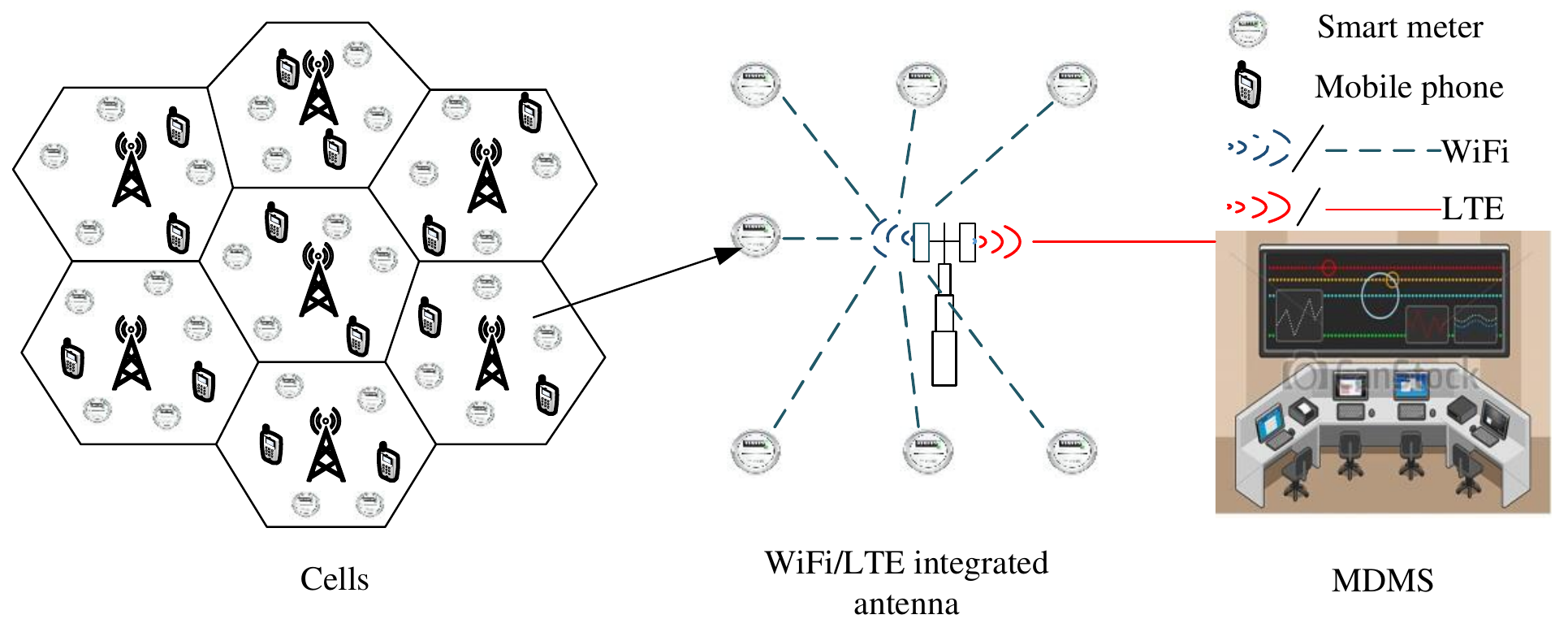}
    \caption{Cell layout for AMI of smart grid.}
    \captionsetup{justification=centering}
    \label{layout}
\end{figure*}

WiFi is a popular communication standard for short range communication. It uses a distributed coordination function (DCF) as its default channel access mode which employs carrier sensing and a four-way handshaking~ \cite{rupasinghe2014licensed}. The DCF mode in WiFi uses the clear channel assessment (CCA) procedure for transmitting the packets. The CCA consists of carrier sensing and energy detection mechanisms to detect the medium whether it is busy or not. So if the level of interference is more than a particular CCA threshold, WiFi nodes will postpone transmission for a random period of time which is called the back-off procedure to avoid packet collision. This possibility may arise due to the transmission of coexisted LTE network. %Therefore, We need appropriate techniques so that LTE and WiFi can access the channel in a fair and good neighboring manner.

The LTE technology, on the other hand, is more systematic and flexible. In case of coexisted LTE and WiFi system in the same band, the primal hindrance is that while WiFi utilizes the OFDM transmission with collision sensed multiple access/collision avoidance (CSMA/CA) protocol, the LTE employs OFDMA channel access technique through which simultaneous low data rate transmission can happen from several UEs with proper frequency and time allocation \cite{chaves2013lte}. Unlike WiFi, LTE does not implement the carrier sensing detection before transmitting the packets. To make the transmission possible it reserves channels for simultaneous transmission.

In typical scenarios of coexistence, the WiFi transmission is most likely to be blocked by LTE transmission. To facilitate co-existence between LTE-U/LAA and WiFi in the same band, mainly three techniques have been proposed in the literature - 1) Dynamic channel selection, 2) Listen Before Talk (LBT) and 3) Co-existence gaps. In \cite{TechRep13}, Qualcom presents an effective channel selection policy depending on interference level. Based on interference at the equipment and network side measured before and during the operation, LTE-U/LAA changes the frequency. In \cite{Ratasuk}, carrier aggregation from licensed  to unlicensed band is proposed with LBT using request-to-send (RTS) and clear-to-send (CTS) prior to original LTE transmission. LBT is mandatory for data offloading in unlicensed band in Europe and Japan. On the other hand, LBT is optional in USA and China market. In \cite{blankspace}, blank subframe in LTE transmission frame is reserved for WiFi transmission while LTE remained silent during this time. Similar method is proposed in \cite{abs} where $n$ of 5 sub-frames of LTE-U/LAA has been kept reserved for WiFi transmission. 

In \cite{rupasinghe2014licensed}, the performance of coexisted system in indoor hot-spot scenario is investigated utilizing a semi-static system level simulator. The results indicated that the performance of WiFi degraded substantially when operated simultaneously with LTE, whereas LTE's performance degraded slightly. Similar results was found in the study of coexistence of LTE and ZigBee \cite{imtiazZigBee}, where ZigBee performance get more effected. In \cite{6916315}, usage of WiFi and LTE is emphasized recommending  WiFi  for high density areas (i.e. urban areas) and LTE for low density areas (i.e. rural areas). In \cite{7436362}, meter data communication with help of a hybrid WiFi/LTE architecture is presented, where WiFi is connected at the bottom layer of LTE. However, the coexistence of WiFi and LTE is not investigated in this study. 

The US federal communications commission (FCC) released 3550-3700 MHz band (also termed as citizen broadband radio service (CBRS)) for shared board band use~\cite{FCC_CBRS_LongReport2,CommLawBlog_2015}. Based on the guideline, the users are grouped into 3 categories: incumbent access (IA) users (tier-1), prioritized access license (PAL) users (tier-2), and general authorized access (GAA) users (tier-3). GAA users has to use the CBRS spectrum providing  privilege to IA and PAL users. In some areas~\cite{NTIA_FastTrack_LongReport} where there will be no IA and PAL activity, 150~MHz can be used by GAA users. On the other hand, at least 80~MHz will be usable for GAA users in vicinity of PAL activity and IA exclusion zone.
This huge band provides free and clean channel for wireless communications such as metering data communication of smart grid~\cite{parvez2016cbrs, parvezaverage, DBLP:journals/corr/abs-1708-09005}. In our study, LTE and WiFi share the CBRS spectrum as GAA users.

In this paper, we proposed LTE and WiFi based AMI for smart grid. In our frame work, smart meters use WiFi to transmit data to data collector/Access point (AP). Data collector collects data from a cluster of meters and send the data to MDMS using LTE. Based on this scenario, we study the performance of coexisted LTE and WiFi in the 3.5 GHz band for AMI communication and usual mobile human-to-human (H2H) communication. We consider a duty cycle based time division duplexing (TDD)-LTE and WiFi for system level simulation on  a collocated network layout. LTE system transmits a fixed duty cycle of a period, and on the other hand, WiFi transmits for the rest of the period. The simulation results demonstrate good neighboring coexistence between LTE and WiFi without significantly hampering each other's performance. Since large amount of clean and free bandwidth is available in CBRS band, coexisted LTE-WiFi based AMI in the CBRS band can be a promising solution for smart grid.

The rest of this paper is organized as follows. In Section~II, system model for the coexistence of LTE/WiFi in 3.5~GHz is presented. Section~III demonstrates the simulation results in the smart scenario. Finally, concluding remarks are presented in Section~IV.

\section {System Model}

\begin{figure*}[t!]
    \centering
    \includegraphics[width= 1\linewidth]{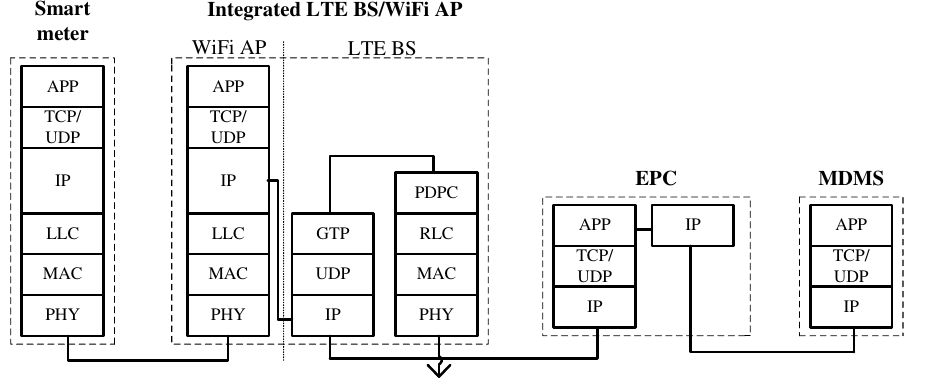}
    \caption{Protocol mapping between different entities of WiFi and LTE network. (AP: access point; EPC= Evolved packet core; APP= application; UDP= user datagram protocol; TCP= transmission control protocol; LLC= logic link control; IP= Internet protocol; PHY= physical; MAC= medium access control; GTP= GPRS tunneling protocol; RLC= radio link control; PDPC= packet Data Convergence Protocol)}
    \captionsetup{justification=centering}
    \label{protocolstack}
\end{figure*}

Let us consider, a collocated LTE-U and WiFi network scenario  where LTE-U and WiFi coexists in the 3.5 GHz band as illustrated in Fig. \ref{layout}. In our proposed framework, smart meters use WiFi  and APs of smart meters use  LTE-U for transferring data. Additionally, collocated WiFi AP and LTE BS are integrated. The data of a cluster of meters is collected by a WiFi AP,  and then  forwarded to integrated LTE BS. Finally LTE BS transmits data to MDMS. The protocol mapping of various entities of WiFi system and LTE network is illustrated in Fig. \ref{protocolstack}. The PHY layer of smart meter is connected with the PHY layer of WiFi AP. On the other hand, The IP layers of WiFi AP and LTE BS are connected in our model. The communication among LTE BS, EPC and MDMS are based on standard LTE system architecture~\cite{5635449}. 

Let us consider, the sets of WiFi APs (i.e. data collector), LTE-U BS, WiFi STAs (i.e. smart meter) and LTE-U UE (i.e. MDMS and other UEs) are given by $S_w$, $S_l$, $U_w^i$ and $U_l^j$ respectively. The transmission power of WiFi AP $i$, LTE BS $j$, meter/WiFi STA $l$ and LTE-U UE/MDMS $m$ are $p_r^i$, $p_r^j$, $p_r^l$ and $p_r^m$.

The channel gain values from WiFi STA/meter  $x$ to WiFi AP $j$, from  LTE UE $a$ to WiFi AP $j$, from LTE-U BS $i$ to WiFi AP $j$ and LTE-U BS $b (i\ne b)$ to WiFi $j$ are $h_{j,r}^x$, $h_{j,r}^a$, $h_{j,r}^i$ and $h_{j,r}^b$ respectively.

The signal-to-noise (SINR) of WiFi AP $j$ during the data reception from meter/WiFi STA $x$ on the resource block $r$ is

\begin{equation}
\mathrm{SINR}_{j,r}^x=\frac{h_{j,r}^x p_r^j}{\sum{h_{j,r}^a p_r^a}+\sum {h_{j,r}^i p_r^i}+\sum {h_{j,r}^b p_r^b}+\sigma^2},
\end{equation}

\noindent where $\sigma^2$ is noise variance. The good SINR value ensures high throughput and low SINR results in reduced throughput performance.\\

The number of successful received bits at WiFi AP $j$ from the WiFi STA $x$, $N_{\mathrm{B}}$   is

\begin{equation}
N_B^x= \mathrm{BT} \sum \log_2 (1+\mathrm{SINR}_{j,r}^x),
\end{equation}

\noindent where B is the bandwidth and 
T is the transmission time such that T=$\sum r$. The number of received bit depends on SINR value.\\

The up link (UL) capacity of WiFi STA/meter $x$ is

\begin{equation}
C^x=\frac{N_B^x}{T_{tx}+T_{wait}},
\end{equation} 

\noindent where $T_{tx}$ and $T_{wait}$ are the transmission and wait time of WiFi, respectively.

For both LTE and WiFi traffic arrival rate  $\lambda$, the distribution function of delay between two packets ($d$) is 
 
\begin{equation}f(d)=\lambda e ^{\lambda d}.\end{equation}

The higher the value of $\lambda$, the more is the number of packet on queue for transmission.

\section{System Parameters and Simulation Results}

To evaluate the performance, a collocated 7 cell architecture is considered as shown in Fig. \ref{layout}. We used the Matlab based simulator which was build based on 3GPP standard and was used in \cite{rupasinghe2014licensed,7506714}. In each cell, for each integrated WiFi AP/LTE-U BS, 10 smart meters (WiFi STAs) and 10 LTE UEs are drooped at random locations. It is noted that one of 10 LTE UEs is to be considered as MDMS. The traffic arrival rates for LTE-U and WiFi are considered as ${\lambda_{LTE}=\lambda_{WiFi}=2.5}$. The PHY and MAC layers of LTE and IEEE 802.11n (WiFi)  are implemented in the simulation environment. In each transmission time interval (TTI), only one UE is scheduled for the DL transmission and the SINR information is sent to the corresponding BS. 

Also based on the number of LTE-U UEs waiting and requesting for the UL transmission during one subframe, bandwidth is equally shared among themselves. The simulation parameter for LTE simulation has been summarized in TABLE \ref{LTETable}. The  parameter value were selected based on 3GPP LTE standard \cite{3GPP}.

\begin{table}[!bp]
	\caption{\ LTE MAC/PHY Parameters.}
	\label{LTETable}
	\centering
	\begin{tabular}{l l}
		\hline 
		\textbf{Parameter} & \textbf{Value} \\
		\hline 
		Frequency band & 3.5 GHz \\
		Bandwidth & 20 MHz \\
		Down link transmission (Tx) Power & 20 dBm \\
		LTE-U UE velocity & 0~ms\\
		Uplink transmission (Tx) Power  & PL Based TPC \\
		Duration of frame  & 10 ms \\
		Scheduling & Round Robin \\
		${P_0}$ & -106 dBm \\
		TTI  & 1 ms \\
		Packet arrival rate $(\lambda )$& 2.5\\
		\hline 
	\end{tabular} 
\end{table}

For WiFi, channel access mechanism CSMA/CA with clear channel assessment assessment (CCA) and enhanced distributed channel access (EDCA) is implemented. WiFi STAs having packets on queue competes for channel access. However, transmission or reception is started only after reception of beacon.  The WiFi STA sends packets when it sense that the channel is idle. Otherwise, the transmission is ceased and the next transmission will be attempted after a random back off period. The WiFi parameter in our simulation are summarized in the TABLE \ref{WiFiTable}. The  parameter value were selected based on study presented in \cite{rupasinghe2014licensed,parvez2016cbrs,7564899}.

\begin{table}[!bp]
	\caption{\ WiFi MAC/PHY Parameters.}
	\label{WiFiTable}
	\centering
	\begin{tabular}{l l}
		\hline 
		\textbf{Parameter} & \textbf{Value} \\ 
		\hline 
		Frequency band& 3.5 GHz \\
		Bandwidth & 20 MHz \\
	    Downlink/Uplink transmission (Tx) power & 23 dBm \\
		WiFi STA/meter velocity & 0~ms\\
		Access category  & Best Effort \\
		MAC protocol  & EDCA \\
		Threshold of CCA sensing & -82 dBm \\
		Threshold of CCA Energy detection  & -65 dBm \\
		Number of service bits in PPDU & 16 bits \\
		Number of tail bits in PPDU & 12 bits \\
		Contention window size  & ${\mathcal{U}\left(0,31\right)}$ \\
		Noise figure  & 6 \\
		Beacon interval & 100 ms \\
		Beacon OFDM symbol detection threshold & 10 dB \\
		Beacon error ratio threshold & 15 \\
		Packet arrival rate $(\lambda )$ & 2.5\\
		
		\hline 
	\end{tabular} 
\end{table}

A physical (PHY) layer abstraction is utilized for shannon capacity calculations of WiFi and LTE-U at the ${4\mu s}$ granularity of WiFi OFDM symbol period of obtaining the number of successfully received bits. FTP Traffic Model-2 \cite{v14} is commonly employed for either WiFi and LTE-U. In our simulation, we used $60\%$ and $80\%$ duty cycle of 50~ms transmission time for LTE. Therefore, WiFi will transmit $40\%$ and $20\%$ duty cycle of the 50~ms period.

\begin{figure*} [!htbp]
 	\centering
 	\subfigure[] { \label{fig_cap_a}\includegraphics[width= 0.45\linewidth,height=5.5 cm]{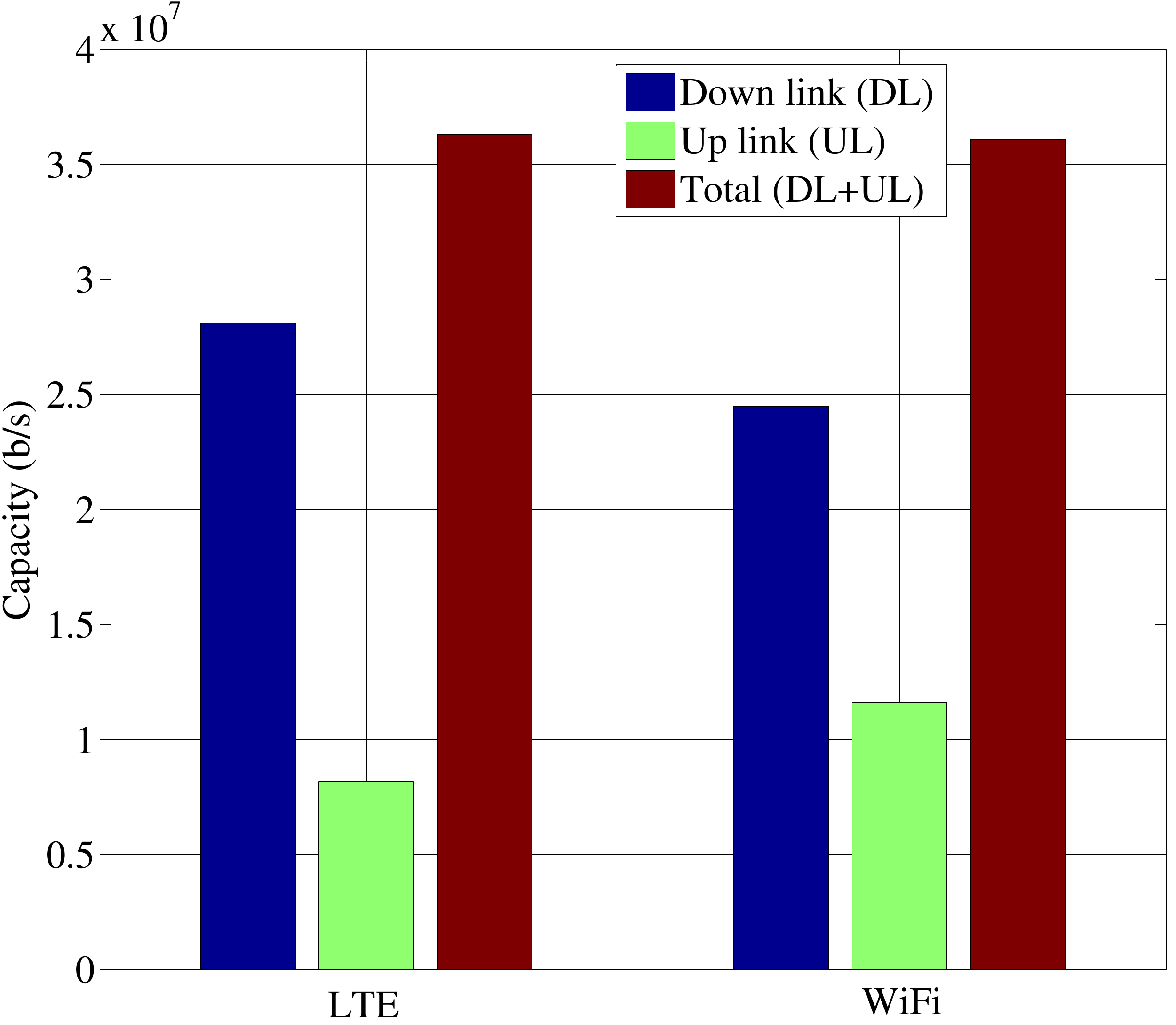}}
 	\subfigure[] {\label{fig_cap_b}\includegraphics[width= 0.45\linewidth,height=5.5 cm]{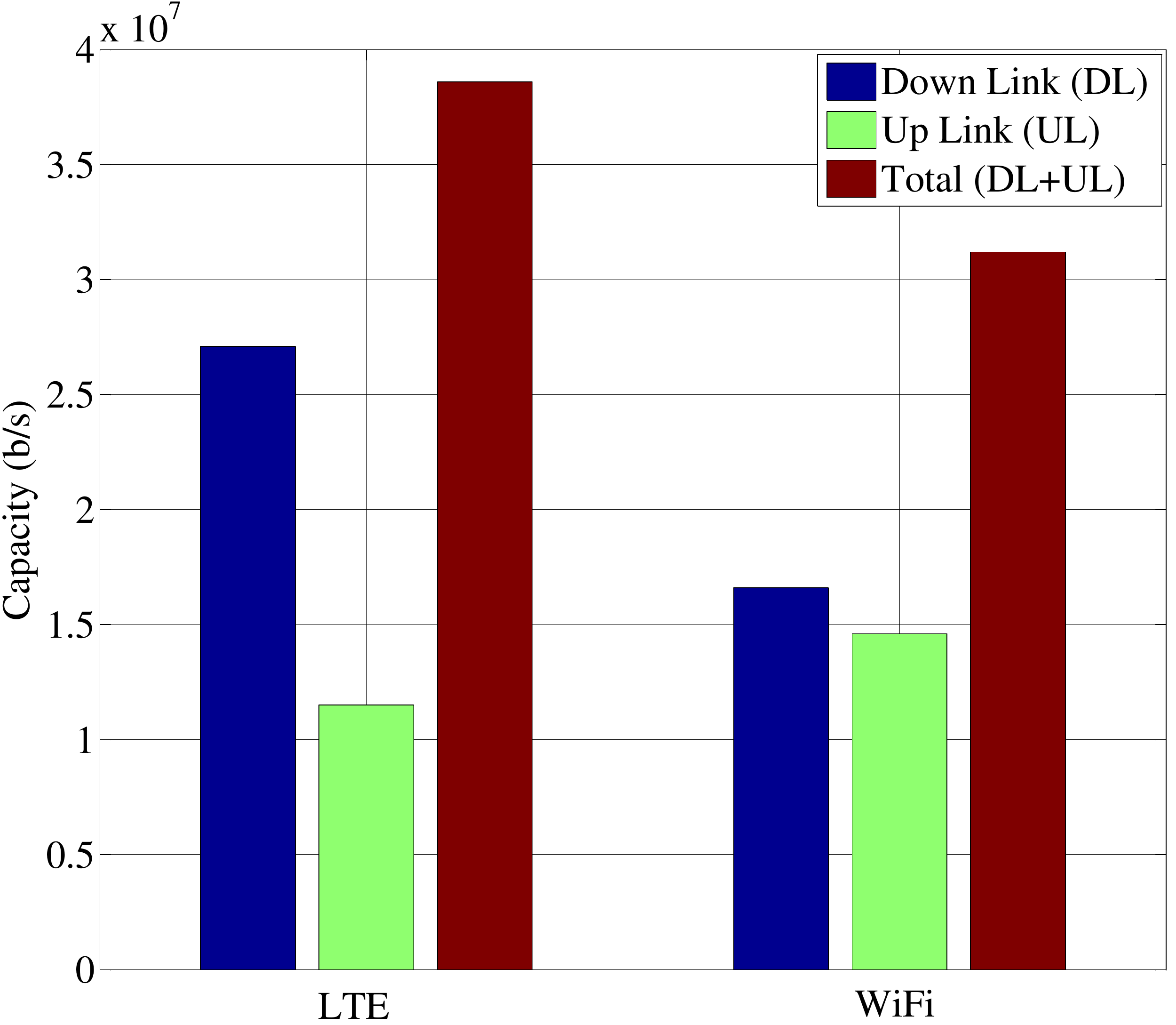}}	
 	\caption{Throughput performance of coexisted LTE/WiFi system (a) Throughput at $60\%$ duty cycle (b) Throughput at $80\%$ duty cycle}
 	\label{fig_capacity}
 	\vspace*{-2mm}
 \end{figure*}

\begin{figure*} [!htbp]
 	\centering
 	\subfigure[] { \label{fig_sinr_a}\includegraphics[width= 0.45\linewidth,height=5.5 cm]{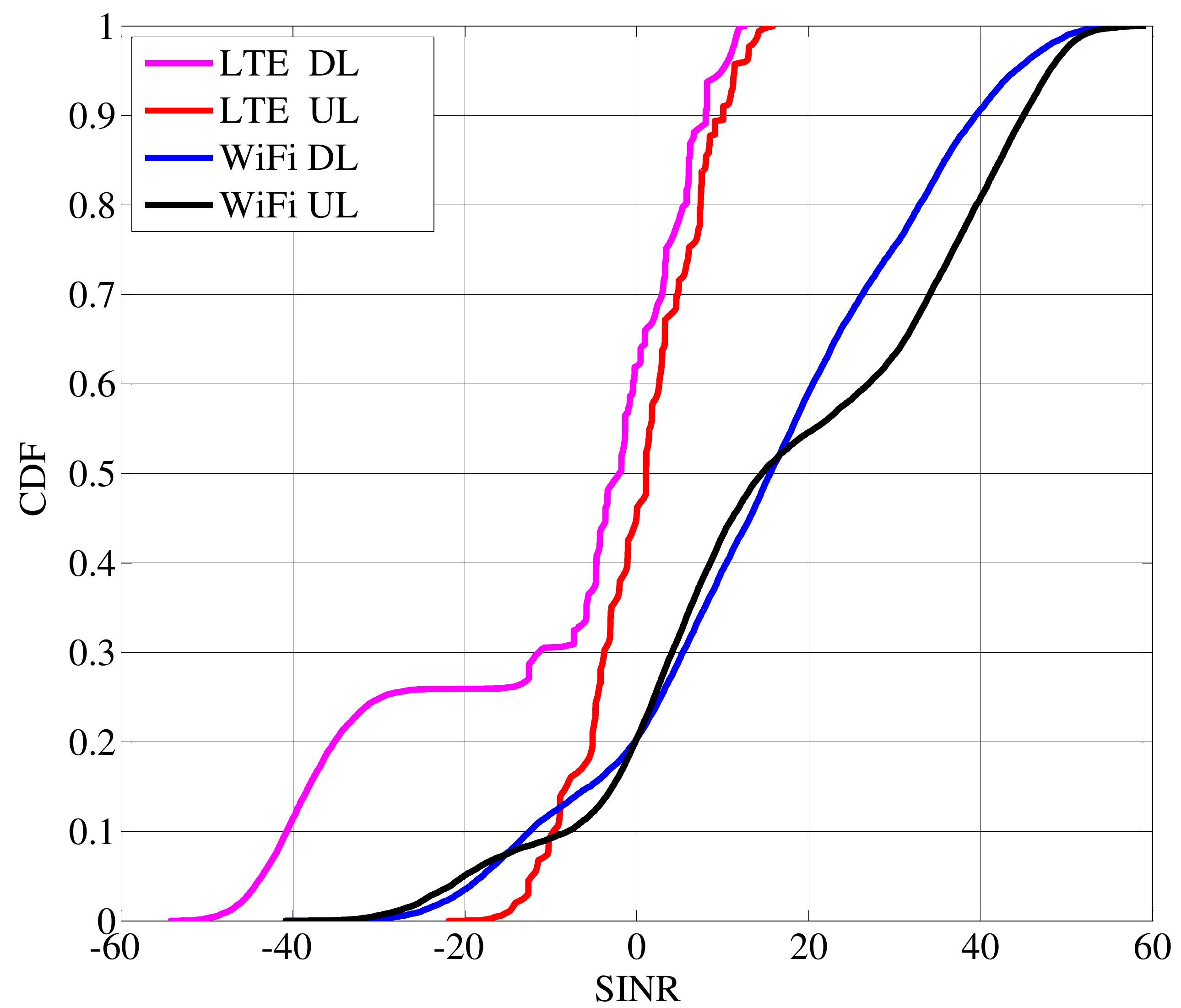}}
 	\subfigure[] {\label{fig_sinr_b}\includegraphics[width= 0.45\linewidth,height=5.5 cm]{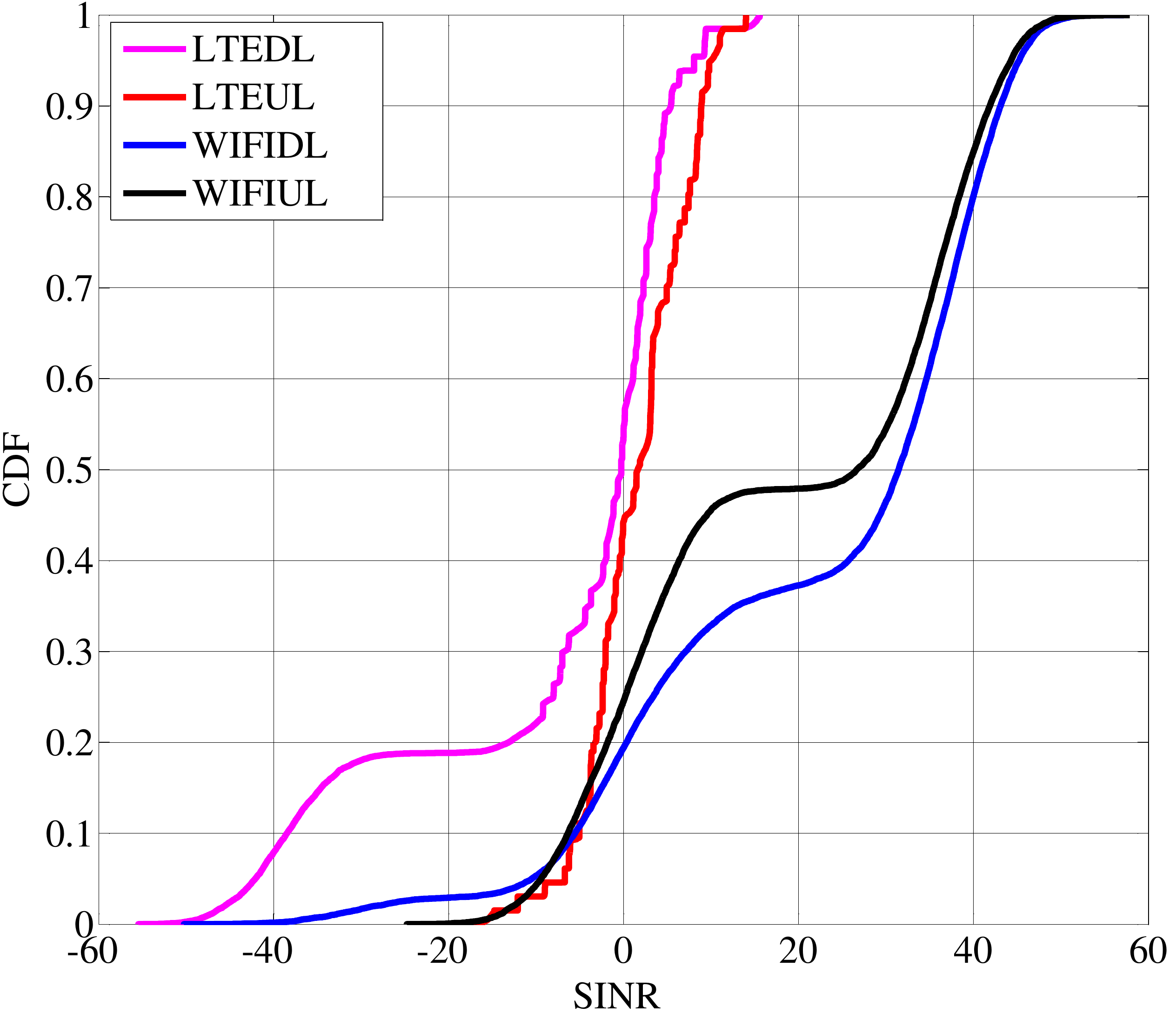}}	
 	\caption{SINR distribution of coexisted LTE/WiFi system (a) SINR distribution at $60\%$ duty cycle (b) SINR distribution at $80\%$ duty cycle}
 	\label{fig:NE1}
 	\vspace*{-2mm}
 \end{figure*}

The throughput performance of coexisted LTE and WiFi in the  smart grid scenario is illustrated in Fig. \ref{fig_capacity}. Referred to Fig. \ref{fig_cap_a}, for $60\%$ duty cycle of LTE-U, the capacity of LTE is 36.3 Mbps and the capacity of WiFi is 36.1~Mpbs. If we increase the duty cycle of LTE-U to $80\%$, the LTE capacity is improved to 38.6~Mbps while the capacity of WiFi is decreased to 31.2~Mbps. This is illustrated in Fig. \ref{fig_cap_b}. The throughput degrade in WiFi is due to the increased transmission backoff on extended transmission time of LTE.

As illustrated in Fig. \ref{fig_sinr_a}, for $60\%$ duty cycle of LTE, the SINR distribution of WiFi is better than that of LTE. However, for $80\%$ duty cycle of LTE, SINR distribution of LTE is improved sightly whereas the SINR distribution of WiFi remained almost same. This is reflected in Fig. \ref{fig_sinr_b}.

The justification of using $60\%$ and $80\%$ duty for LTE is that LTE will be used not only for meter data communication to MDMS, but also it will be used for human-to-human communication (i.e. personal mobile communication). Therefore, we provide more access to LTE transmission. However, more time (i.e. duty cycle) can be allocated for WiFi transmission based on the number of smart meters.

\section{Conclusion}
In this paper, we proposed a LTE and WiFi based metering infrastructure in the 3.5~GHz band. In our architecture, meter uses WiFi whereas AP uses LTE for transferring data. LTE transmits for  a fixed duty cycle of a period, whereas WiFi transmits in the rest of the period. However, the duty cycle can be manipulated based on the number of smart meters. The promising simulation results demonstrate that good neighborhood spectrum sharing can be possible without harming each other's performance. Moreover, the 3.5~GHz band has large clean and free bandwidth for data communication. Therefore, 3.5~GHz band sharing by LTE and WiFi can be a promising candidate communication architecture for metering infrastructure of smart grid.
 
\bibliographystyle{IEEEtran}
\bibliography{IEEEabrv,bibolte}

\end{document}